\documentclass[aps,prd,showpacs,preprint]{revtex4}

\usepackage{amsmath}
\usepackage{dcolumn}
\usepackage{graphicx}

\newcommand{\XSB}{{\chi\text{SB}}}
\newcommand{\XQM}{{\chi\text{QM}}}

\begin{document}
\title{Understanding the Flavor Symmetry Breaking and Nucleon
  Flavor-Spin Structure within Chiral Quark Model} 

\author{Zhan Shu}
\affiliation{Department of Physics, Peking University, Beijing 100871,
  China} 

\author{Xiao-Lin Chen} \affiliation{Department of Physics,
  Peking University, Beijing 100871, China} 

\author{Wei-Zhen Deng} 
\email{dwz@th.phy.pku.edu.cn}
\affiliation{Department of Physics, Peking University, Beijing 100871,
  China} 


\begin{abstract}
  In $\XQM$, a quark can emit Goldstone bosons. The flavor symmetry
  breaking in the Goldstone boson emission process is used to intepret
  the nucleon flavor-spin structure.  In this paper, we study the
  inner structure of constituent quarks implied in $\XQM$ caused by
  the Goldstone boson emission process in nucleon.  From a simplified
  model Hamiltonian derived from $\XQM$, the intrinsic wave functions
  of constituent quarks are determined. Then the obtained transition
  probabilities of the emission of Goldstone boson from a quark can
  give a reasonable interpretation to the flavor symmetry breaking in
  nucleon flavor-spin structure.
\end{abstract}
\maketitle

\section{INTRODUCTION}

The measurements of the polarized structure functions of the nucleon
in deep inelastic scattering(DIS)
experiments\cite{dis-1,dis-2,dis-3,dis-4} show the complication in
proton spin structure. Only a portion of the proton spin is carried by
valence quarks. Moreover, several
experiments\cite{asym-1,asym-2,asym-3} clearly indicate the
$\bar{u}$-$\bar{d}$ asymmetry as well as the existence of the strange
quark content $\bar{s}$ in the proton sea.  Also the distribution of
strange quark in the proton sea is polarized negative.
The DIS results deviate significantly from the na\"ive quark
model (NQM) expectation.

NQM gives many fairly good descriptions of hadron properties. Why does
NQM work? It is a puzzle that the quarks inside a hadron could be
treated as non-relativistic particles in NQM.  The chiral quark model
($\chi$QM) tries to bridge between QCD and NQM.  It was originated by
Weinberg\cite{cqm-1} and formulated by Manohar and Georgi\cite{cqm-2}.
Between the QCD confinement scale ($\Lambda_{\text{QCD}}\simeq$200MeV)
and a chiral symmetry breaking scale ($\Lambda_{\XSB}\simeq$1GeV), the
strong interaction is described by an effective Lagrangian of quarks
$q$, gluons $g$ and Numbu-Goldstone bosons $\Pi$.  An important
feature of the $\chi$QM is that, betweetn $\Lambda_{\text{QCD}}$ and
$\Lambda_{\XSB}$, the internal gluon effects in a hadron can be small
compared to the internal Goldstone bosons $\Pi$ and quarks $q$, so the
effective degrees of freedom in this region can be $q$ and $\Pi$.

It is interesting that $\chi$QM can also be used to explain why NQM
does not work in the above DIS experiments. By the emission of
Goldstone boson, $\chi$QM allows the fluctuation of a quark $q$ into a
recoiling quark plus a Goldstone boson $q\to q'\Pi$ . The $q'\Pi$ system
then further splits to generate quark sea through
\begin{itemize}
\item the helicity-flipping process
  \begin{equation}\label{pro1}
    q_{\uparrow}\longrightarrow
    \Pi+q_{\downarrow}^{\prime}\longrightarrow
    (q\bar{q^{\prime}})+q_{\downarrow}^{\prime}
  \end{equation}
\item and the helicity-non-flipping process
  \begin{equation}\label{pro2}
    q_{\uparrow}\longrightarrow 
    \Pi+q_{\uparrow}^{\prime}\longrightarrow
    (q\bar{q^{\prime}})+q_{\uparrow}^{\prime}
  \end{equation}
\end{itemize}
where the subscript indicates the helicity of quark. In both the
process, $q^{\prime}\Pi$ is in a relative P-wave state. In the
helicity-flipping process (\ref{pro1}), the orbital angular momentum
along helicity direction must be $\langle l_{z}\rangle=+1$. In the
helicity-non-flipping process (\ref{pro2}), $\langle l_{z}\rangle=0$.
The process cause a modification of the spin content of the nucleon
because a quark changes its helicity in (\ref{pro1}). Also it causes a
modification of the flavor content because the generated quark sea
from $\Pi$ is flavor dependent\cite{Eichten,ChengLi}.

$\chi$QM was first used to explain the nucleon sea flavor asymmetry
and the smallness of the quark spin fraction by Eichten, Hinchliffe
and Quigg\cite{Eichten}. The flavor asymmetry of sea quark
distribution arises from the mass differences in different quark
flavors and in different Goldstone bosons.  Only the lightest Goldstone
Boson $\pi$ was considered since its contribution dominates. From
a perturbation calculation, the probability for an up quark
to emit a $\pi^+$ was estimated to be $a=0.083$.  This would induce a
flavor asymmetry in parton distributions of nucleon and other hadrons.

However, the estimated transition probability is not enough to full
account the flavor asymmetry in DIS experiments.  Contribution from
other $\Pi$'s and even $\eta^{\prime}$ was considered by Cheng and
Li\cite{ChengLi}.  Explicit $SU_f(3)$ breaking in the transition
probabilities was later introduced in refs.~\onlinecite{ChengLi2,Song}
and further used by several
authors\cite{ChengLi1,Linde,Gupta1,Gupta,Gupta2,Yu}.  Nevertheless, in
all these calculations, the transition probabilities were put into
model by hand.  To fit the experimental data, the probability of an up
quark emitting $\pi^+$ needs to be set to $a\gtrsim 0.1$, which is
about $20\%$ larger than the perturbation calculation. Although the
probability of $\pi$ emission can be enlarged by using a higher
momentum cut off $\Lambda>\Lambda_{\XQM}$ in the perturbation
calculation \cite{Pirner}, however, the chiral quark model is no
longer valid at arbitrary high energies $\Lambda \gg \Lambda_{\XQM}$.

We should not be surprised by this discrepancy since the $\chi$QM
works in a region right above the QCD confinement scale
$\Lambda_{\text{QCD}}$.  There one may expect the confinement effect
is important and the perturbative calculation of QCD may contain large
error.  However, there is another essential difference between the above
model calculations and the perturbation calcultion. In the
perturbation calculation, the emitted Goldstone bosons are virtual
particles. In the above model calculations which are closely related to
NQM, however, the Goldstone bosons are close to mass shell under the
non-relativistic approximation.

Since $\chi$QM can be a bridge between NQM and QCD, it is interesting to
explore $\chi$QM from NQM side where we use the wave function method. This
will give the above model calculations a concrete foundation in NQM and
help us further understand the flavor symmetry breaking mechanism.

In this paper, we will use wave function method to investigate the
flavor symmetry breaking in $\chi$QM.  In a conventional quark
model\cite{Godfrey}, a hadron consists of confined constituent quarks
and its wave function is constructed in the configuration space of the
constituent quarks. To incorporate the transition process of 
emitting Goldstone boson of $\chi$QM into the quark model, the constituent
quarks will have intrinsic wave functions within the configuration
$q+q'\Pi$.

In Sec.~II, we first present the composite wave function of
constituent quarks including components of $q'\Pi$.  The wave
functions and the transition probabilities of $q\to q'\Pi$ are
determined from a simplified $\chi$QM Hamiltionian.  In Sec.~III and
Sec.~IV, the obtained transition probabilities are used to calculate
nucleon flavor-spin structure and baryon octet magnetic moments
respectively. The numerical results and a brief summary are presented
in Sec.~V.

\section{The Wave Function of a Constituent quark}

In $\chi$QM, the effective Lagrangian below the chiral symmetry
breaking scale $\Lambda_\XQM$ involves quarks, gluons, and Goldstone
bosons. The first few terms in this Lagrangian are\cite{cqm-2}:
\begin{align}
  \mathcal{L}_{\XQM}&=\bar{\psi}(i D_{\mu}+V_{\mu})\gamma^{\mu}\psi
  +ig_{A}\bar{\psi}A_{\mu}\gamma^{\mu}\gamma^{5}\psi \notag\\
  &-m\bar{\psi}\psi+\frac14 f_\pi^2 \text{tr} \partial^\mu \Sigma^\dag
  \partial_\mu\Sigma +...
\end{align}
where $D_{\mu}=\partial_{\mu}+igG_{\mu}$ is the gauge-covariant
derivative of QCD, $G_{\mu}$ the gluon field and $g$ the strong
coupling constant. The dimensionless axial-vector coupling
$g_{A}=0.7524$ is determined from the axial charge of the nucleon. $m$
represents the constituent quark masses due to chiral symmetry
breaking. The pseudoscalar decay constant is $f_\pi\approx93$MeV.  The
$\Sigma$ field, vector currents $V_\mu$ and axial-vector currents
$A_\mu$ are given in terms of the Goldstone boson fields $\Phi$
\begin{align}
  \Phi &=
  \begin{bmatrix}
    \frac{1}{\sqrt{2}}\pi^{0}+\frac{1}{\sqrt{6}}\eta & \pi^{+} & K^{+} \\
    \pi^{-} & -\frac{1}{\sqrt{2}}\pi^{0}+\frac{1}{\sqrt{6}}\eta & K^{0} \\
    K^{-} & \bar{K^{0}} & -\frac{2}{\sqrt{6}}\eta 
  \end{bmatrix}, \\
  \Sigma &=\exp(i \frac{\sqrt2\Phi}{f_\pi}), \\
  \begin{pmatrix}
    V_{\mu} \\
    A_{\mu} 
  \end{pmatrix}
  &=\frac{1}{2}(\xi^{\dagger}\partial_{\mu}\xi\pm\xi\partial_{\mu}\xi^{\dagger}),
  \\
  \xi&=\exp(i\frac{\Phi}{\sqrt{2}f_\pi}).
\end{align}
An expansion of the currents in powers of $\Phi/f_\pi$ yields the
effective interaction between $\Pi$ and $q$\cite{Eichten}
\begin{equation}\label{Lint}
  \mathcal{L}_{I}=-\frac{g_{A}}
  {\sqrt{2}f_\pi}\bar{\psi}\partial_{\mu}\Phi\gamma^{\mu}\gamma_{5}\psi.
\end{equation}
This allows the fluctuation of a quark into a recoil quark
plus a Goldstone boson $q \to q'\Pi$. 

In quark model, a hadron is built with constituent quarks. In
accordance with $\chi$QM, we should treat a constituent quark as a
composite particle including such components $q'\Pi$.  Here we denote
the wave function of a composite constituent quark as
$|q\rangle\rangle$.  At rest,
\begin{equation}\label{cqf}
  |q\rangle\rangle = z^q |q\rangle + \sum_{q'\Pi} x^q_{q'\Pi} 
  |q'\Pi\rangle.
\end{equation}
In our paper, the state normalization relation is always taken as
\begin{equation}
  \langle p | p' \rangle = \delta^3(\mathbf{p}-\mathbf{p}').
\end{equation}

The above wave function is of essential importance in our work. The
square of the modulus of the coefficient of each $q'\Pi$
configuration is just the probability for the corresponding $\Pi$
emission process
\begin{equation}
  P_{q\to q'\Pi}=|x_{q'\Pi}^q|^2,
\end{equation}
and
\[
|z^q|^2=(1-\sum_{q'\Pi} P_{q\to q'\Pi})
\]
is the probability of no $\Pi$ emission.

To determine the wave function (\ref{cqf}), we first construct a
simplified Hamiltonian in the degrees of freedom $q$ and $\Pi$,
\begin{equation}\label{hcq}
H=H_{0}+H_{B}+H_{I}.
\end{equation}
$H_{0}$ represents the kinetic energies of $q$ and $\Pi$. 
It reads
\begin{equation}
  H_0 = \int d^3x \left\{\bar\psi(i\alpha\cdot\nabla+m)\psi
  +\frac12 \text{Tr}[{\dot\Phi}^2+(\nabla\Phi)^2]
  +\frac12 \sum_\Pi m_\Pi^2 (\Phi^\Pi)^2 \right\},
\end{equation}
where $m_\Pi$ is the physical mass of $\Pi$ which is nonzero and
nondegenerate.  
\begin{equation}
  H_I = - \int d^3x \mathcal{L}_I,
\end{equation}
is the $\XQM$ interaction. $H_B$ is an accessary interaction which is
needed to bind the $q'\Pi$ together. In our simplified Hamiltonian, we
will not disscuss the explicit formalism of $H_B$. Instead, we will
put some physical restriction conditions on it later in this section, which
is sufficient to our calculation.

From $H_0$, we can expand free fields $\psi$ and $\Pi$ in terms of
creation and annihilation operators
\begin{align}
  \label{qfree}
  \psi^q(x) &= \int \frac{d^3p}{(2\pi)^{3/2}} \frac1{\sqrt{2E_{\mathbf{p}}^q}}
  \sum_s \left[ a^q_{\mathbf{p}s} u^q(\mathbf{p},s) e^{-i p\cdot x}
    +b^{q\dag}_{\mathbf{p}s}(t) v^q(\mathbf{p},s) e^{i p\cdot x}
  \right]_{p^0=E^q_{\mathbf{p}}}, \\
  \label{Phifree}
  \Phi^\Pi(x) &= \int \frac{d^3p}{(2\pi)^{3/2}} 
  \frac1{\sqrt{2E_{\mathbf{p}}^\Pi}}
  \left[ c^\Pi_{\mathbf{p}} e^{-i p\cdot x} + c^{\Pi\dag}_{\mathbf{p}}
    e^{i p\cdot x} \right]_{p^0=E^\Pi_{\mathbf{p}}},
\end{align}
where 
\[
E_{\mathbf{p}}^q=\sqrt{\mathbf{p}^{2}+m_q^2}
\]
is the quark energy of flavor $q$,
\[
E_{\mathbf{p}}^\Pi=\sqrt{\mathbf{p}^{2}+m_\Pi^2}
\]
is the energy of Goldstone boson $\Pi$. $a_{\mathbf{p}s}^{q\dag}$ and
$b_{\mathbf{p}r}^{q\dag}$ are the creation operators of quark $q$ and
anti-quark $\bar{q}$
\begin{equation}\label{etcr1}
  \{a^q_{\mathbf{p}r},a_{\mathbf{p}'s}^{q\dag}\}
  =\{b^q_{\mathbf{p}r},b_{\mathbf{p}'s}^{q\dag}\}
  =\delta^{(3)}(\mathbf{p}-\mathbf{p}')\delta_{rs}.
\end{equation}
$c_{\mathbf{p}}^{\Pi\dag}$ is the creation operator of $\Pi$
\begin{equation}\label{etcr2}
  [c^\Pi_{\mathbf{p}},c_{\mathbf{p}'}^{\Pi\dag}]
  =\delta^{(3)}(\mathbf{p}-\mathbf{p}').
\end{equation}

Next, we will replace the field $\psi$ and $\Phi$ in the Hamiltonian
(\ref{hcq}) with the free field of (\ref{qfree}) and
(\ref{Phifree}). Then we can express the Hamiltonian in creation and
annihilation operators, for example
\begin{equation}\label{freeenergy}
  H_{0}=\sum_{q}\;\sum_{s}\int
  d^3p\;E_{\mathbf{p}}^{q} [a_{\mathbf{p}s}^{q\dag}a_{\mathbf{p}s}^q+
  b_{\mathbf{p}s}^{q\dag}b_{\mathbf{p}s}^{q}]+\sum_{\Pi}\int
  d^3p\;E_{\mathbf{p}}^\Pi c_{\mathbf{p}}^{\Pi\dag}c_{\mathbf{p}}^\Pi.
\end{equation}

In all the model calculations \cite{ChengLi,ChengLi2,ChengLi1,Linde,
  Song,Gupta1,Gupta,Gupta2,Yu}, the emitted $\Pi$ is assumed bound to
the quark source.  To represent that $q'\Pi$ are bound, we use the
well known SHO function as their spatial wave function
\begin{align}\label{final}
  |q\Pi\rangle&=\frac{1}{\sqrt{N}}(-i) \int d^3p 
  |\mathbf{p}| e^{-\frac{p^2}{2\lambda^2}}\;[Y_{1}(\theta,\phi)
  \;c_{-\mathbf{p}}^{\Pi\dag}\;a_{\mathbf{p}}^{q\dag}]_{1/2} \;|0\rangle,
  \\
  |q\Pi\uparrow\rangle&=\frac{1}{\sqrt{N}}\sqrt{\frac23}(-i) \int d^3p 
  |\mathbf{p}| e^{-\frac{p^2}{2\lambda^2}}\;Y_{11}(\theta,\phi)
  \;c_{-\mathbf{p}}^{\Pi\dag}\;a_{\mathbf{p}\downarrow}^{q\dag} \;|0\rangle
  \notag\\
  &-\frac{1}{\sqrt{N}}\sqrt{\frac13}(-i) \int d^3p 
  |\mathbf{p}| e^{-\frac{p^2}{2\lambda^2}}\;Y_{10}(\theta,\phi)
  \;c_{-\mathbf{p}}^{\Pi\dag}\;a_{\mathbf{p}\uparrow}^{q\dag} \;|0\rangle,
\end{align}
where $\lambda$ is the ``characteristic radius'' parameter in Gaussian
function.  $1/\sqrt{N}$ is the normalization factor,
\begin{equation}
  N=\int dp\;p^4\;e^{-\frac{p^2}{\lambda^2}}=\frac38 \sqrt{\pi}\lambda^5.
\end{equation}

However, we need a binding interaction $H_B$ in the Hamiltonian.  Yet
we do not know how to write out the explicit form of $H_B$. However,
$H_B$ should provide enough binding energy.  That is, for the $q'\Pi$
system, we must have
\begin{equation}
\langle q\Pi|H_0 + H_B|q\Pi \rangle \le m_q + m_\Pi.
\end{equation}
That is
\begin{equation}
E_B= \langle q\Pi|H_B|q\Pi \rangle \le m_q + m_\Pi 
-\langle q\Pi|H_0|q\Pi \rangle = m_q-E^q+m_\Pi-E^\Pi.
\end{equation}
As a rough estimation, we will take the mininum value of $E_B$
\begin{equation}
E_B= -\max_{q,\Pi}\{E^q-m_q+E^\Pi-m_\Pi\} = -(E^u-m_u+E^\pi-m_\pi).
\end{equation}

Then the wave function of a composite constituent quark is determined
from Schr\"odinger equation
\begin{equation}
  H|q\rangle\rangle = M_q|q\rangle\rangle.
\end{equation}
After taking the above simplification, we need only solve a matrix
eigen-value problem
\begin{equation}
\label{hcqp}
\begin{pmatrix} a & B \\ B^T & C
\end{pmatrix} 
\begin{pmatrix} z^q\\X^q
\end{pmatrix} 
= M_q
\begin{pmatrix} z^q\\X^q
\end{pmatrix}, 
\end{equation}
where
\begin{align*}
  a\delta^3(0) &= \langle q|H|q\rangle, \\
  B_{q'\Pi}\delta^3(0) &= \langle q|H|q'\Pi\rangle, \\
  C_{q'\Pi;q''\Pi'}\delta^3(0) &= \langle q'\Pi|H|q''\Pi'\rangle, \\
  X^q_{q'\Pi} &= x^q_{q'\Pi}.
\end{align*}

For example, let us consider the process $u$ emitting $\Pi$.
There are four possible $|q'\Pi\rangle$ states generated by the
fluctuations of a $u$ quark: 
$|u\pi^{0}\rangle$, $|u\eta\rangle$,
$|d\pi^{+}\rangle$ and $|sK^{+}\rangle$. Thus 
\begin{equation}\label{cqf1}
  |u\rangle\rangle=z^u|u\rangle
  +x_{u\pi^0}^u|u\pi^{0}\rangle
  +x_{u\eta}^u|u\eta\rangle+x_{d\pi^+}^u|d\pi^{+}\rangle
  +x_{sK^+}^u|sK^{+}\rangle.
\end{equation}
Taking these wave functions as basis, we can calculate the matrix of
the Hamiltonian in (\ref{hcqp}). 
\begin{equation}
  a=m_{u}.
\end{equation}
$C$ is diagonalized. Its diagonal matrix elements are calculated from
$H_0$
\begin{align}
C_{u\pi^{0};u\pi^{0}}&=\frac1N \int
dp\;p^4\;e^{-\frac{p^{2}}{\lambda^{2}}}(\sqrt{\mathbf{p}^{2}+m_{u}^{2}}
+\sqrt{\mathbf{p}+m_{\pi^{0}}^{2}}) +E_B,\\
C_{u\eta;u\eta}&=\frac1N \int
dp\;p^4\;e^{-\frac{p^{2}}{\lambda^{2}}}(\sqrt{\mathbf{p}^{2}+m_{u}^{2}}
+\sqrt{\mathbf{p}+m_{\eta}^{2}}) +E_B,\\
C_{d\pi^{+};d\pi^{+}}&=\frac1N \int
dp\;p^4\;e^{-\frac{p^{2}}{\lambda^{2}}}(\sqrt{\mathbf{p}^{2}+m_{d}^{2}}
+\sqrt{\mathbf{p}+m_{\pi^{+}}^{2}}) +E_B,\\
C_{sK^{+};sK^{+}}&=\frac1N \int
dp\;p^4\;e^{-\frac{p^{2}}{\lambda^{2}}}(\sqrt{\mathbf{p}^{2}+m_{s}^{2}}
+\sqrt{\mathbf{p}+m_{K^{+}}^{2}}) +E_B.
\end{align}
$B$ is calculated from $H_I$
\begin{align}
B_{u\pi^{0}} &=-\frac{g_A}{2\sqrt2\pi f_\pi}\frac1{\sqrt{N}}\int
dp\;p^4 e^{-\frac{p^{2}}{2\lambda^{2}}}
\frac{1}{\sqrt{4E_{\mathbf{p}}^{u}E_{-\mathbf{p}}^{\pi^{0}}}}
\cdot \sqrt{E_{\mathbf{p}}^{u}+m_u}
\left(1+\frac{E_{-\mathbf{p}}^{\pi^{0}}}{E_{\mathbf{p}}^{u}+m_u}\right), \\
B_{u\eta} &=-\frac{g_A}{2\sqrt6\pi f_\pi}\frac1{\sqrt{N}}\int
dp\;p^4 e^{-\frac{p^{2}}{2\lambda^{2}}}
\frac{1}{\sqrt{4E_{\mathbf{p}}^{u}E_{-\mathbf{p}}^{\eta}}}
\cdot \sqrt{E_{\mathbf{p}}^{u}+m_u}
\left(1+\frac{E_{-\mathbf{p}}^{\eta}}{E_{\mathbf{p}}^{u}+m_u}\right), \\
B_{d\pi^{+}} &=-\frac{g_A}{2\pi f_\pi}\frac1{\sqrt{N}}\int
dp\;p^4 e^{-\frac{p^{2}}{2\lambda^{2}}}
\frac{1}{\sqrt{4E_{\mathbf{p}}^{d}E_{-\mathbf{p}}^{\pi^{+}}}}
\cdot \sqrt{E_{\mathbf{p}}^{d}+m_d}
\left(1+\frac{E_{-\mathbf{p}}^{\pi^{+}}}{E_{\mathbf{p}}^{d}+m_d}\right), \\
B_{sK^+} &=-\frac{g_A}{2\pi f_\pi}\frac1{\sqrt{N}}\int
dp\;p^4 e^{-\frac{p^{2}}{2\lambda^{2}}}
\frac{1}{\sqrt{4E_{\mathbf{p}}^{s}E_{-\mathbf{p}}^{K^+}}}
\cdot\sqrt{E_{\mathbf{p}}^{s}+m_s}
\left(1+\frac{E_{-\mathbf{p}}^{K^{+}}}{E_{\mathbf{p}}^{s}+m_s}\right) .
\end{align}
By diagonalizing this Hamiltonian matrix, we will obtain a new mass of
the constituent $u$ quark $M_u$ and its composite wave function.  The
constituent masses and wave functions of $d$ and $s$ quarks can be
obtained similarly. We have
\begin{align}\label{cqf2}
  |d\rangle\rangle&=z^d|d\rangle
  +x_{d\pi^0}^d|d\pi^{0}\rangle
  +x_{d\eta}^d|d\eta\rangle+x_{u\pi^-}^d|u\pi^{-}\rangle
  +x_{sK^0}^d|sK^{0}\rangle, \\
  \label{cqf3}
  |s\rangle\rangle&=z^s|s\rangle
  +x_{s\eta}^s|s\eta\rangle
  +x_{d\bar{K}^0}^s|d\bar{K}^0\rangle+x_{uK^-}^s|uK^-\rangle.
\end{align}
From isospin symmetry, $m_u=m_d$, we have
\begin{equation}
z^d=z^u;\quad x^d_{d\pi^0}=-x^u_{u\pi^0};\quad 
x^d_{u\pi^-} = x^u_{d\pi^+}; \quad ...
\end{equation}
However, since $m_u \ne m_s$, one should notice that
\begin{equation}
z^s\ne z^u;\quad x^s_{d\bar{K}^0} \ne x^d_{sK^0}; \quad
x^s_{uK^-} \ne x^u_{sK^+}.
\end{equation}

After the diagonalization, the Goldstone bosons $\Pi$ are separated
from quarks $q$ approximately. With only degrees of freedom $q$ one can
rebuild the quark model and so $M_u$, $M_d$, $M_s$ should be
regarded as the constituent quark masses in quark model.

\section{FLAVOR AND SPIN STRUCTURE OF PROTON}\label{sec3}

Having known the wave functions of constituent quark $q$ and the
transition amplitudes of $q$ emitting each Goldstone bosons $\Pi$, we are
able to calculate the quark distribution in a constituent quark
following refs.~\onlinecite{ChengLi,ChengLi2,Song}. In $\XQM$, $\Pi$
will further split into a quark-antiquark pair. By substituting the
quark contents of $\Pi$ into wave functions (\ref{cqf1}), (\ref{cqf2})
and (\ref{cqf3}), we can rewrite the wave functions of constituent
quark $q$ as
\begin{align}
|u\rangle\rangle&=z^u|u\rangle
+\left(\frac{x^u_{u\eta}}{\sqrt{6}}+\frac{x^u_{u\pi^0}}{\sqrt{2}}\right)
|u(u\bar{u})\rangle
+\left(\frac{x^u_{u\eta}}{\sqrt{6}}-\frac{x^u_{u\pi^0}}{\sqrt{2}}\right)
|u(d\bar{d})\rangle \notag\\
&-\frac{2x^u_{u\eta}}{\sqrt{6}}|u(s\bar{s})\rangle
+x^u_{d\pi^+}|d(u\bar{d})\rangle+x^u_{sK^+}|s(u\bar{s})\rangle, \\
|d\rangle\rangle&=z^u|d\rangle
+\left(\frac{x^u_{u\eta}}{\sqrt{6}}-\frac{x^u_{u\pi^0}}{\sqrt{2}}\right)
|d(u\bar{u})\rangle
+\left(\frac{x^u_{u\eta}}{\sqrt{6}}+\frac{x^u_{u\pi^0}}{\sqrt{2}}\right)
|d(d\bar{d})\rangle \notag\\
&-\frac{2x^u_{u\eta}}{\sqrt{6}}|d(s\bar{s})\rangle
+x^u_{d\pi^+}|u(d\bar{u})\rangle+x^u_{sK^+}|s(d\bar{s})\rangle, \\
|s\rangle\rangle&=z^s|s\rangle
+\frac{x^s_{s\eta}}{\sqrt{6}}|s(u\bar{u})\rangle
+\frac{x^s_{s\eta}}{\sqrt{6}}|s(d\bar{d})\rangle
-\frac{2x^s_{s\eta}}{\sqrt{6}}|s(s\bar{s})\rangle \notag\\
&+x^s_{d\bar{K}^0}|d(s\bar{d})\rangle+x^s_{uK^-}|u(s\bar{u})\rangle.
\end{align}

Then the antiquark and quark flavor contents of the proton ($uud$) are
\begin{align}
\bar{u}&=2\left|\frac{x^u_{u\eta}}{\sqrt{6}}
  +\frac{x^u_{u\pi^0}}{\sqrt{2}}\right|^{2}
+\left|\frac{x^u_{u\eta}}{\sqrt{6}}-\frac{x^u_{u\pi^0}}{\sqrt{2}}\right|^{2}
+|x^u_{d\pi^+}|^{2}, &
u&=\bar{u}+2, \\
\bar{d}&=\left|\frac{x^u_{u\eta}}{\sqrt{6}}
  +\frac{x^u_{u\pi^0}}{\sqrt{2}}\right|^{2}
+2\left|\frac{x^u_{u\eta}}{\sqrt{6}}-\frac{x^u_{u\pi^0}}{\sqrt{2}}\right|^{2}
+2|x^u_{d\pi^+}|^{2}, &
d&=\bar{d}+1, \\
\bar{s}&=2|x^u_{u\eta}|^{2}+3|x^u_{sK^+}|^{2}, &
s&=\bar{s}.
\end{align}

Some important quantities depending on the above quark distribution
are: the Gottfried sum rule
$I_{G}=\frac{1}{3}+\frac{2}{3}(\bar{u}-\bar{d})$ whose deviation
indicates the $\bar{u}$-$\bar{d}$ asymmetry in proton sea;
$\bar{u}/\bar{d}$ measured through the ratio of muon pair production
cross sections; and the fractions of quark flavors in proton
$f_{q}=\frac{q+\bar{q}}{\Sigma(q+\bar{q})}$, $f_{3}=f_{u}-f_{d}$ and
$f_{8}=f_{u}+f_{d}-2f_{s}$.

We can further calculate the spin structure of proton. Here one should
consider the effects of configuration mixing generated by spin-spin
forces\cite{Rujula}. We take the baryon wave functions from the quark
model calculation\cite{Isgur,Koniuk,Isgur1}.  The proton wave function
for example, is expressed as
\begin{equation}{\label{baryonstate}}
  \left|P,{\frac{1}{2}}^{+}\right\rangle
  =0.90|P_{8}^{2}S_{S}\rangle-0.34|P_{8}^{2}S^{\prime}_{S}\rangle
  -0.27|P_{8}^{2}S_{M}\rangle
\end{equation}
where the baryon $SU(6)\otimes O(3)$ wave functions are denoted as
$|B_{N}^{2S+1}L_{\sigma}\rangle$, $N$ is $SU(3)$ multiplicity. $S$,
$L$ are the total spin and total orbital angular momentum while
$\sigma=S,M,A$ denotes the permutation symmetry of $SU(6)$.  The spin
polarization functions will be remarkably affected by configuration
mixing. Following refs.~\onlinecite{Linde,Gupta}, we define the number
operator by
\[
\hat{N}=n_{u\uparrow}u_{\uparrow} + n_{u\downarrow}u_{\downarrow}
+n_{d\uparrow}d_{\uparrow} + n_{d\downarrow}d_{\downarrow}
+n_{s\uparrow}s_{\uparrow} + n_{s\downarrow}s_{\downarrow},
\]
where $n_{q\uparrow}$, $n_{q\downarrow}$ are the number of
$q_{\uparrow}$, $q_{\downarrow}$ quarks. The spin structure of the
``mixed'' proton is given by
\begin{align}\label{mixedproton}
  \hat P &\equiv \left\langle
    P,{\frac{1}{2}}^+ \right|N \left|P,{\frac{1}{2}}^+\right\rangle \notag\\
  &=(0.90^{2}+0.34^{2})\left(\frac{5}{3}u_{\uparrow}+\frac{1}{3}u_{\downarrow}
    +\frac{1}{3}d_{\uparrow}+\frac{2}{3}d_{\downarrow}\right)
  +0.27^{2}\left(\frac{4}{3}u_{\uparrow}
    +\frac{2}{3}u_{\downarrow}+\frac{2}{3}d_{\uparrow}
    +\frac{1}{3}d_{\downarrow}\right).
\end{align}
The spin structure after considering
$\Pi$-emission is obtained by replacing for every quark in
eq.~(\ref{mixedproton}) by
\begin{equation}
  q_{\uparrow,\downarrow}\longrightarrow(1-\Sigma P_i)q_{\uparrow,\downarrow}
  +P_{flipping}(q_{\uparrow,\downarrow})
  +P_{non-flipping}(q_{\uparrow,\downarrow}),
\end{equation}
where $P_{flipping}(q_{\uparrow,\downarrow})$ and
$P_{non-flipping}(q_{\uparrow,\downarrow})|$ are the
probabilities of quark helicity flipping and non-flipping for
$q_{\uparrow,\downarrow}$ respectively. For example, in the case of
$u_\uparrow$ quark we have,
\begin{equation*}
  P_{flipping}(u_{\uparrow})=\frac23 \left[
  (|x^u_{u\pi^0}|^{2}+|x^u_{u\eta}|^{2})u_{\downarrow}
  +|x^u_{d\pi^+}|^{2}d_{\downarrow}+|x^u_{sK^+}|^{2}s_{\downarrow}
  \right],
\end{equation*}
and
\begin{equation*}
  P_{non-flipping}(u_{\uparrow})=\frac13 \left[
    (|x^u_{u\pi^0}|^{2}+|x^u_{u\eta}|^{2})u_{\uparrow}
    +|x^u_{d\pi^+}|^{2}d_{\uparrow}+|x^u_{sK^+}|^{2}s_{\uparrow}
  \right].
\end{equation*}
Finally the spin polarization functions defined as $\Delta
q=q_{\uparrow}-q_{\downarrow}$ are
\begin{align}
  \Delta u &=(0.90^{2}+0.34^{2})\left[\frac{4}{3}
    -\left(\frac{114|x^u_{u\pi^0}|^{2}+48|x^u_{u\eta}|^{2}
        +36|x^u_{sK^+}|^{2}}{27}\right)\right] \notag\\
  &+0.27^{2}\left[\frac{2}{3}-\left(\frac{66|x^u_{u\pi^0}|^{2}
        +24|x^u_{u\eta}|^{2}+18|x^u_{sK^+}|^{2}}{27}\right)\right], \\
  \Delta d &=(0.90^{2}+0.34^{2})\left[-\frac{1}{3}
    +\left(\frac{6|x^u_{u\pi^0}|^{2}+12|x^u_{u\eta}|^{2}
        +9|x^u_{sK^+}|^{2}}{27}\right)\right] \notag\\
  &+0.27^{2}\left[\frac{1}{3}-\left(\frac{42|x^u_{u\pi^0}|^{2}
        +12|x^u_{u\eta}|^{2}+9|x^u_{sK^+}|^{2}}{27}\right)\right], \\
  \Delta s &=-\frac{|x^u_{sK^+}|^{2}}{3}.
\end{align}

There are several measured quantities which can be expressed in
terms of the above spin  polarization functions. The quantities
usually calculated are $\Delta_3=\Delta u-\Delta d$ and
$\Delta_8=\Delta u+\Delta d-2\Delta s$, obtained from the neutron
$\beta$-decay and the weak decays of hyperons respectively.
Another important quantity is the flavor singlet component of the
total quark spin content defined as $2 \Delta \Sigma=\Delta
u+\Delta d+\Delta s\,$. We also calculate some weak axial-vector
form factors which are also related to the spin polarization
functions, $(G_{A}/G_{V})_{\Lambda\rightarrow
p}=\frac{1}{3}(2\Delta u-\Delta d-\Delta s)$,
$(G_{A}/G_{V})_{\Sigma^{-}\rightarrow n}=\Delta d-\Delta s$, and
$(G_{A}/G_{V})_{\Xi^{-}\rightarrow\Lambda}=\frac{1}{3}(\Delta
u+\Delta d-2\Delta s)$.

\section{BARYON OCTET MAGNETIC MOMENTS}\label{sec4}

Considering the relative angular momentum between quark and Goldstone
boson $\Pi$, the magnetic moment operator of a $q\Pi$ system is
\begin{align}\label{mmo}
  \hat{\mathbf{\mu}}_{q\Pi} &=\frac{e_{q}}{m_{q}}\;\hat{\mathbf{s}}
  +\frac{e_{q}}{2\sqrt{{\mathbf{p}^{2}_{q}}
      +{m^{2}_{q}}}}\frac{\sqrt{{\mathbf{p}^{2}_{\Pi}}
      +{m^{2}_{\Pi}}}}{\sqrt{{\mathbf{p}^{2}_{q}}+{m^{2}_{q}}}
    +\sqrt{{\mathbf{p}^{2}_{\Pi}}+{m^{2}_{\Pi}}}}\;\hat{\mathbf{l}}\notag\\
  &+\frac{e_{\Pi}}{2\sqrt{{\mathbf{p}^{2}_{\Pi}}+{m^{2}_{\Pi}}}}
  \frac{\sqrt{{\mathbf{p}^{2}_{q}}+{m^{2}_{q}}}}
  {\sqrt{{\mathbf{p}^{2}_{q}}+{m^{2}_{q}}}
    +\sqrt{{\mathbf{p}^{2}_{\Pi}}+{m^{2}_{\Pi}}}}\;\hat{\mathbf{l}}
\end{align}
where $e_{q}$ and $e_{\Pi}$ are the electric charges carried by $q$
and $\Pi$ respectively, $\hat{\mathbf{s}}$ the quark spin operator and
$\hat{\mathbf{l}}$ the relative angular momentum bewteen $q$ and
$\Pi$.  The first term in Eq(\ref{mmo}) is the intrinsic magnetic
moment of quark and the other two terms are the contribution of the
orbital angular momentum.  Here we have to consider the relativistic
effect since the relative momentum of $q$ or $\Pi$ are
comparable to their masses in the $q\Pi$ system
\[
\mathbf{p}_{q,\Pi} \sim \Lambda \sim m_{q,\Pi}.
\]

With the SHO wave functions of (\ref{final}), the magnetic moment of
$q\Pi$ system (\ref{mmo}) can be readily calculated. Then we can recalculate
the magnetic moments of constituent quarks taking into account of the
relativistic effect. For example, the magnetic moments of the $u$
quark is
\begin{align}\label{mmu}
\mu_{u}&=|z^u|^{2}\langle
u_{\uparrow}|\hat{\mathbf{\mu}}|u_{\uparrow}\rangle +P_{u\to u\pi^0}\langle
u\pi^{0}|\hat{\mathbf{\mu}}|u\pi^{0}\rangle+P_{u\to u\eta}\langle
u\eta|\hat{\mathbf{\mu}}|u\eta\rangle \notag \\
&+P_{u\to d\pi^+}\langle
d\pi^{+}|\hat{\mathbf{\mu}}|d\pi^{+}\rangle+P_{u\to sK^+}\langle
sK^{+}|\hat{\mathbf{\mu}}|sK^{+}\rangle,
\end{align}
where
\begin{equation}
\langle
u_{\uparrow}|\hat{\mathbf{\mu}}|u_{\uparrow}\rangle=\frac{e_{u}}{2m_{u}},
\end{equation}
and the contribution from $q\Pi$ systems are
\begin{align}
\langle
u\pi^{0}|\hat{\mathbf{\mu}}|u\pi^{0}\rangle&=-\frac{e_{u}}{6m_{u}}+
\frac{e_{u}}{3N} \int
dp\;\frac{\sqrt{p^2+m^2_{\pi}}}{\sqrt{p^2+m^2_{u}}
+\sqrt{p^2+m^2_{\pi}}}\frac{1}{\sqrt{p^2+m^2_{u}}}p^4\;
e^{-\frac{p^{2}}{\lambda^{2}}} ,\\
\langle
u\eta|\hat{\mathbf{\mu}}|u\eta\rangle&=-\frac{e_{u}}{6m_{u}}+
\frac{e_{u}}{3N} \int
dp\;\frac{\sqrt{p^2+m^2_{\eta}}}{\sqrt{p^2+m^2_{u}}+\sqrt{p^2+m^2_{\eta}}}
\frac{1}{\sqrt{p^2+m^2_{u}}}p^4\;e^{-\frac{p^{2}}{\lambda^{2}}},
\\
\langle
d\pi^{+}|\hat{\mathbf{\mu}}|d\pi^{+}\rangle&=-\frac{e_{d}}{6m_{d}}+
\frac{e_{d}}{3N} \int
dp\;\frac{\sqrt{p^2+m^2_{\pi}}}{\sqrt{p^2+m^2_{d}}+\sqrt{p^2+m^2_{\pi}}}
\frac{1}{\sqrt{p^2+m^2_{d}}}p^4\;e^{-\frac{p^{2}}{\lambda^{2}}} \notag\\
&+\frac{e_{\pi^{+}}}{3N} \int
dp\;\frac{\sqrt{p^2+m^2_{d}}}{\sqrt{p^2+m^2_{d}}+\sqrt{p^2+m^2_{\pi}}}
\frac{1}{\sqrt{p^2+m^2_{\pi}}}p^{4}\;e^{-\frac{p^{2}}{\lambda^{2}}},
\\
\langle
sK^{+}|\hat{\mathbf{\mu}}|sK^{+}\rangle&=-\frac{e_{s}}{6m_{s}}+
\frac{e_{s}}{3N} \int
dp\;\frac{\sqrt{p^2+m^2_{K}}}{\sqrt{p^2+m^2_{s}}+\sqrt{p^2+m^2_{K}}}
\frac{1}{\sqrt{p^2+m^2_{s}}}p^4\;e^{-\frac{p^{2}}{\lambda^{2}}} \notag\\
&+ \frac{e_{K^{+}}}{3N} \int
dp\;\frac{\sqrt{p^2+m^2_{s}}}{\sqrt{p^2+m^2_{s}}+\sqrt{p^2+m^2_{K}}}
\frac{1}{\sqrt{p^2+m^2_{K}}}p^4\;e^{-\frac{p^{2}}{\lambda^{2}}}.
\end{align}
The magnetic moments of $d$ and $s$ quarks can be calculated
similarly. 

One can easily obtain the octet baryon magnetic moments by replacing
the valence quarks inside the baryons with the corresponding
constituent quarks. Again we take proton as an example,
\begin{equation}
  \mu_{p}=(0.90^{2}+0.34^{2})\left(\frac{4}{3}\mu_{u}-\frac{1}{3}\mu_{d}\right)
  +0.27^{2}\left(\frac{2}{3}\mu_{u}+\frac{1}{3}\mu_{d}\right).
\end{equation}
If we replace the $\mu_q$ by (\ref{mmu}), $\mu_p$ can be further
expressed as the baryon magnetic moment in conventional quark model
plus the contribution from the Goldstone boson emission process
\cite{Franklin}. The magnetic moments for other octet baryons can be
calculated similarly.

\section{NUMERICAL RESULTS AND CONCLUSIONS}\label{sec5}

In the numerical calculation, most of the parameters can be taken from the
experimental data or the chiral quark model.  We collect these fixed input
parameters of our calculation in Table \ref{table1}.  Here we have used the
the physical masses of Goldstone bosons\cite{pdg}.
\begin{table}[h]
  \caption{The fixed input parameters from chiral quark model and 
    experimental data.\label{table1}}
  \begin{ruledtabular}
    \begin{tabular}{c c c c c}
      $g_{A}$ & $f_\pi$(MeV) & $m_{\pi}$(MeV) &$m_{K}$(MeV)&$m_{\eta}$(MeV)\\
      \hline
      0.7524 & 93 & 135  & 494 & 548 \\
    \end{tabular}
  \end{ruledtabular}
\end{table}

For the quark masses, since our work focuses on the inner context of the
constituent quarks in quark model, naturally we will refer to the
quark masses from quark model, instead of the chiral quark model
values. Here we will use the quark mass values from the widely
accepted Isgur's quark model\cite{Godfrey} as shown in Table
\ref{table1.5}. However, one should be cautious that, in our model,
it is the quark with the Goldstone boson mixing which corresponds to
the constituent quark in quark model.  That is, mass values $M_q$ after
the diagonalization process should be set to the quark masses in
Isgur's model.  Our strategy is to adjust the quark masses $m_q$ in
the model Hamiltonian to fit the $M_q$ values.

Finally we are left only with one free parameter $\lambda$ which
describes the confinement of the emitted Goldstone boson in our
model. An overall fit to the experimental data of nucleon flavor-spin
structure and octet baryon magnetic moments shows that the best value
should be $\lambda$=152MeV. With this value of $\lambda$ and a minimun
binding energe $E_B=-218$MeV, the ``bare'' values of quark masses
$m_q$ without Goldstone boson mixing are shown also in Table
\ref{table1.5}.

\begin{table}[h]
  \caption{The quark masses with vs. without Goldstone boson mixing.
    \label{table1.5}}
  \begin{ruledtabular}
    \begin{tabular}{c c c c c c}
      $\lambda$ & $E_B$(MeV) & $m_{u,d}$(MeV) & 
      $m_s$(MeV) &$M_{u,d}$(MeV)&$M_s$(MeV)\\
      \hline
      152 & $-$218 & 288 & 474  & 220 & 419 \\
    \end{tabular}
  \end{ruledtabular}
\end{table}

Transition probabilities of the light and strange quarks to various
$q^{\prime}\Pi$ systems are given in Table \ref{table2} and
\ref{table3} respectively. The probability of a $u$ quark emitting a
$\pi^{+}$ $P(u\rightarrow d+\pi^{+})$=0.145 is significantly larger
than the perturbation calculation $a$=0.083. One may notice that the
$\lambda$ parameter value $152$MeV in our wave function, which is
below $\Lambda_{\text{QCD}}$, is rather small than another energy
scale $\Lambda_{\XQM}$ in chiral quark model. Surely this will weaken the
interaction between $q$ and $q'\Pi$. However, one should also notice
that the binding energy $E_B=-218$MeV will make the energy of a $q\Pi$
system much close to the single quark energy. This will enhance the mixing
of $q'\Pi$ components in a constituent quark.

Also, we notice that the asymmetry between the probabilities of
$u(d)\to s+K$ and $s\to u(d)+\bar{K}$. Whether this asymmetry leads to
any observable consequence in hadron structure needs further
investigation.

\begin{table}[h]
\caption{Transition probabilities of a $u$ quark to various
$q^{\prime}\Pi$ systems and the mass of constituent $u$
quark.\label{table2}}
\begin{ruledtabular}
\begin{tabular}{c c c c c c}
  $u\to u+\pi^{0}$ & $u\to u+\eta$  
  & $u\to d+\pi^{+}$ & $u\to s+K^{+}$& no GB-emission & $M_{u}$\\
  \hline
   0.072 & 0.003 & 0.145 & 0.010 & 0.770 & 220MeV \\
\end{tabular}
\end{ruledtabular}
\end{table}
\begin{table}[h]
\caption{Transition probabilities of a $s$ quark to various
$q^{\prime}\Pi$ systems and the mass of constituent $s$
quark.\label{table3}}
\begin{ruledtabular}
\begin{tabular}{c c c c c}
  $s\to s+\eta $  & $s\to u+K^{-}$ 
  & $s\to d+\bar{K^{0}}$& no GB-emission & $M_{s}$ \\
  \hline
   0.012 & 0.071 & 0.071 & 0.846  & 419MeV \\
\end{tabular}
\end{ruledtabular}
\end{table}

Next, we will compare our calculate results with the experimental data.
Since our emphasis is on the substructure of a constituent quark in
NQM, here we also quote the results from NQM.  In Table \ref{table4},
the calculated flavor and spin structures of the proton are shown. It
should be mentioned that the quark spin polarization functions can be
further corrected by the gluon
anomaly\cite{Altarelli,Carlitz,Efremov,Linde,Gupta,Song} through
\begin{equation}
\Delta q(Q^{2})=\Delta q-\frac{\alpha_{s}(Q^{2})}{2\pi}\Delta g(Q^{2}),
\end{equation}
and the flavor singlet component of the total helicity is
modified accordingly as
\begin{equation}
\Delta\Sigma(Q^{2})=\Delta\Sigma-\frac{3\alpha_{s}(Q^{2})}{4\pi}\Delta g(Q^{2}),
\end{equation}
where $\Delta q(Q^{2})$ and $\Delta\Sigma(Q^{2})$ are the
experimentally measured quantities, $\Delta q$ and $\Delta\Sigma$
correspond to the calculated quantities without gluon
correction. Using the experimental data
$\Sigma(Q^{2}=5\text{GeV}^2)=0.19\pm0.02$\cite{dis-2},
$\alpha_{s}(Q^{2}=5\text{GeV}^{2})=0.285\pm0.013$\cite{pdg}, and our
result $\Delta\Sigma$=0.346, the gluon polarization $\Delta
g(Q^{2})$ is estimated to be 2.293. Both the results with and without gluon
polarization corrections are presented in Table~\ref{table4}. 
The inclusion of gluon polarization leads to a better agreement with
experimental data for the spin structure.

\begin{table}[ht]
\caption{The calculated values for the quark flavor distribution
functions and spin polarization functions in proton, as compared
with experimental data and NQM results.\label{table4}}
\begin{ruledtabular}
\begin{tabular}{c  c  c   c  c}
 & Data & NQM &  \multicolumn{2}{c}{Our Model} \\
 \cline{4-5}  
 &   &   &   With $\Delta g$ & Without $\Delta g$\\
 \hline
 $\Delta u$ & $0.85\pm0.05$\cite{dis-2} & $1.33$ & $0.864$ & $0.968$\\
 $\Delta d$ & $-0.41\pm0.05$\cite{dis-2} & $-0.33$   & $-0.377$& $-0.274$\\
 $\Delta s$ & $-0.07\pm0.05$\cite{dis-2} & $0$  & $-0.107$& $-0.003$\\
 $\Delta_{3}=(G_{A}/G_{V})_{n\rightarrow p}$ & $1.270\pm0.003$\cite{pdg} 
 & $1.67$ & $1.242$& $1.242$\\
 $(G_{A}/G_{V})_{\Lambda\rightarrow p}$ & $0.718\pm0.015$\cite{pdg} 
 & $1$ & $0.737$& $0.737$\\
 $(G_{A}/G_{V})_{\Sigma\rightarrow n}$ & $-0.340\pm0.017$\cite{pdg} 
 & $-0.33$ & $-0.270$& $-0.270$\\
 $(G_{A}/G_{V})_{\Xi\rightarrow \Lambda}$ & $0.25\pm0.05$\cite{pdg} 
 & $0.33$ & $0.234$& $0.234$\\
 $\Delta_{8}$ & $0.58\pm0.025$\cite{dis-2} 
 & $1$ & $0.701$& $0.701$\\
 $\Delta\Sigma$ & $0.19\pm0.02$\cite{dis-2} 
 & $0.5$ &$0.190$&$0.346$\\
 \hline
 $\bar{u}$ & $-$ &    &\multicolumn{2}{c}{$0.264$}\\
 $\bar{d}$ & $-$ &    &\multicolumn{2}{c}{$0.392$}\\
 $\bar{s}$ & $-$ &    &\multicolumn{2}{c}{$0.036$}\\
 $\bar{u}-\bar{d}$ & $-0.118\pm0.015$\cite{asym-2} & $0$  
 & \multicolumn{2}{c}{$-0.128$}\\
 $\bar{u}/\bar{d}$ & $0.67\pm0.06$\cite{asym-2} & $1$ 
 &\multicolumn{2}{c}{$0.674$}\\
 $I_{G}$ & $0.254\pm0.005$\cite{asym-2} & $0.33$ 
 &\multicolumn{2}{c}{$0.248$}\\
 $f_{u}$ & $-$ &     &\multicolumn{2}{c}{$0.577$}\\
 $f_{d}$ & $-$ &     &\multicolumn{2}{c}{$0.407$}\\
 $f_{s}$ & $0.10\pm0.06$\cite{Grasser} &  $0$  
 &\multicolumn{2}{c}{$0.017$}\\
 $f_{3}$ & $-$ &     &\multicolumn{2}{c}{$0.170$}\\
 $f_{8}$ & $-$ &     &\multicolumn{2}{c}{$0.950$}\\
 $f_{3}/f_{8}$ & $0.21\pm0.05$\cite{ChengLi1} & $0.33$ 
 &\multicolumn{2}{c}{$0.179$}\\
\end{tabular}
\end{ruledtabular}
\end{table}

The calculated magnetic moments of octet baryons are given in Table
\ref{table5}.  Although the deviation is somewhat around $30\%$ in the
case of $\Xi^{-}$, our overall fit to octet baryon magnetic moments is
in good agreement with experiments.  Also it should be mentioned that
even in the case of $\Xi^{-}$ the fit can perhaps be improved if
corrections due to pion loops are taken into account\cite{Theberge,Cohen}.

\begin{table}[h]
\caption{The caculated octet baryon magnetic moments in nuclear
magneton, as compared with experiments and the results of
NQM.\label{table5}}
\begin{ruledtabular}
\begin{tabular}{c  c  c   c}
  Octet baryons & Data\cite{pdg} & NQM\cite{NQM} 
  &   Our model \\
  \hline
  $p$ & $2.79\pm0.00$ &  2.72  &$2.73$\\
  $n$ & $-1.91\pm0.00$ &  -1.81   &$-1.91$\\
  $\Sigma^{-}$ & $-1.16\pm0.025$ &  -1.01   &$-1.23$\\
  $\Sigma^{+}$ & $2.46\pm0.01$ & $2.61$  &$2.67$\\
  $\Xi^{0}$ & $-1.25\pm0.0014$ & $-1.41$  &$-1.36$\\
  $\Xi^{-}$ & $-0.65\pm0.002$ & $-0.50$  &$-0.44$\\
  $\Lambda$ & $-0.61\pm0.004$ &   $-0.59$  &$-0.56$\\
  $\Sigma\Lambda$ & $1.61\pm0.08$ &  $1.51$  &$1.63$\\
\end{tabular}
\end{ruledtabular}
\end{table}

In the model calculations \cite{ChengLi,ChengLi2,ChengLi1,Linde,
  Song,Gupta1,Gupta,Gupta2,Yu}, the Goldstone boson sector in $\XQM$
is usually extended to include the $\eta'$ meson with $U(3)$ symmetry.
According to Cheng and Li\cite{ChengLi}, in the large $N_{c}$ limit of
QCD, there are nine Goldstone bosons including the usual octet and the
singlet $\eta^{\prime}$. Thus an constituent quark can also transit to
a quark-$\eta^{\prime}$ system. We have also made an $U(3)$
calculation.  With the inclusion of $\eta^{\prime}$, we find that the
probabilities for $\eta^{\prime}$-emission from light and strange
quarks $P(u\rightarrow u+\eta^{\prime})$=$P(d\rightarrow
d+\eta^{\prime})$=0.0021 and $P(s\rightarrow s+\eta^{\prime})$=0.0018
which are negligibly small as compared to those of octet Goldstone
boson emissions. We therefore conclude that the contribution of
$\eta^{\prime}$ is not important, due to the obvious axial $U(1)$
symmetry breaking in meson mass spectra
$m_{\eta^{\prime}}>m_{K,\eta}$.

To summarize, the $\chi$QM builds a bridge between the QCD and
low-energy quark model. This allows us to understand the mechanism of
flavor symmetry breaking and nucleon flavor-spin structure in NQM
through the consideration of the sea quark and Goldstone bosons in the
substructure of constituent quarks.  Using the simple SHO wave
function, we have modeled the wave functions of the composite
constituent quarks and thus estimated the transition probabilities for
Goldstone boson emissions. These transition probabilities indeed
reflect the flavor SU(3) symmetry breaking in $\chi$QM from the
differences in quark masses $m_{s}>m_{u,d}$ and differences in
Goldstone bosons masses $m_{K,\eta}>m_{\pi}$ and they are roughly in agreement
with the parametrizations of other model calculations
\cite{ChengLi,ChengLi2,ChengLi1,Linde, Song,Gupta1,Gupta,Gupta2,Yu}.
The fit to both the flavor-spin structure of nucleon and octet baryon
magnetic moments are in good agreement with experiments.

\begin{acknowledgments}
Zhan Shu would like to thank Fan-Yong Zou and Yan-Rui Liu for useful
discussions. This work was supported by the National Natural Science
Foundation of China under Grants 10675008.
\end{acknowledgments}


\begin{thebibliography}{99}
\bibitem{dis-1} J.~Ashman et~al. (European Muon), 
  Phys. Lett. \textbf{B206}, 364 (1988); 
  Nucl. Phys. \textbf{B328}, 1 (1990).
\bibitem{dis-2} B.~Adeva et~al. (Spin Muon), Phys. Lett. \textbf{B302}, 
  533 (1993); 
  P.~Adams et~al. (Spin Muon), Phys. Rev. \textbf{D56}, 5330 (1997).
\bibitem{dis-3} P.~L.~Anthony et~al. (E142), Phys. Rev. Lett. \textbf{71}, 
  959 (1993).
\bibitem{dis-4} K.~Abe et~al. (E143), Phys. Rev. Lett. \textbf{74}, 346 (1995).

\bibitem{asym-1} P.~Amaudruz et~al. (New Muon),
  Phys. Rev. Lett. \textbf{66}, 2712 (1991); 
  M.~Arneodo et~al. (New Muon), Phys. Rev. \textbf{D50}, R1 (1994).
\bibitem{asym-2} E.~A.~Hawker et~al. (E866/NuSea),
  Phys. Rev. Lett. \textbf{80}, 3715 (1998); 
  J.~C.~Peng et~al. (E866/NuSea), Phys. Rev. \textbf{D58}, 092004 (1998); 
  R.~S.~Towell et~al. (E866/NuSea), \emph{ibid}. \textbf{D64}, 052002 (2001).
\bibitem{asym-3} A.~Baldit et~al. (NA51), Phys. Lett. \textbf{B332}, 244 (1994).

\bibitem{cqm-1} S.~Weinberg, Physica \textbf{A96}, 327 (1979).
\bibitem{cqm-2} A.~Manohar and H.~Georgi, Nucl. Phys. \textbf{B234}, 327
  (1984).

\bibitem{Eichten} E.~J.~Eichten, I.~Hinchliffe, and C.~Quigg, Phys.
  Rev. \textbf{D45}, 2269 (1992).

\bibitem{ChengLi} T.~P.~Cheng and L.-F.~Li, Phys. Rev. Lett.
  \textbf{74}, 2872 (1995).

\bibitem{ChengLi2} T.~P.~Cheng and L.-F.~Li, Phys. Rev. \textbf{D57},
  344 (1998).
\bibitem{Song} X.~Song, J.~S.~McCarthy, and H.~J.~Weber, Phys. Rev.
  \textbf{D55}, 2624 (1997); 
  X.~Song, \emph{ibid.}, \textbf{D57}, 4114 (1998).

\bibitem{ChengLi1} T.~P.~Cheng and L.-F.~Li, Phys. Rev. Lett.
  \textbf{80}, 2789 (1998).

\bibitem{Linde} J.~Linde, T.~Ohlsson, and H.~Snellman, Phys. Rev.
  \textbf{D57}, 452 (1998);
  T.~Ohlsson and H.~Snellman, Eur. Phys. J. \textbf{C7}, 501 (1999). 
\bibitem{Gupta1} H.~Dahiya and M.~Gupta, Phys. Rev. \textbf{D64}, 014013
  (2001).
\bibitem{Gupta} H.~Dahiya and M.~Gupta, Phys. Rev. \textbf{D66}, 051501(R)
  (2002);
  \textbf{D67}, 114015 (2003). 
\bibitem{Gupta2} H.~Dahiya, M.~Gupta, and J.~M.~S.~Rana, Int. J. Mod.
  Phys. \textbf{A21}, 4255 (2006).
\bibitem{Yu} L.~Yu, X.-L.~Chen, W.-Z. Deng, and S.-L. Zhu, Phys. Rev. 
  \textbf{D73}, 114001 (2006).

\bibitem{Pirner} S.~Baumgartner, H.~J.~Pirner, K.~C.~Konigsmann, and
  B.~Povh, Z. Phys. \textbf{A353}, 397 (1996).

\bibitem{Godfrey} S.~Godfrey and N.~Isgur, Phys. Rev \textbf{D32}, 189 (1985); 
  S.~Capstick and N.~Isgur, Phys. Rev. \textbf{D34}, 2809 (1986).
\bibitem{Rujula} A.~De~Rujula, H.~Georgi, and S.~L.~Glashow, Phys. Rev.
  \textbf{D12}, 147 (1975).
\bibitem{Isgur} N.~Isgur and G.~Karl, Phys. Rev. \textbf{D18}, 4187 (1978).
\bibitem{Koniuk} R.~Koniuk and N.~Isgur, Phys. Rev. \textbf{D21}, 1868 (1980)
\bibitem{Isgur1} N.~Isgur, G.~Karl, and R.~Koniuk,
  Phys. Rev. Lett. \textbf{41}, 1269 (1978); 
  N.~Isgur and G.~Karl, Phys. Rev.  \textbf{D21}, 3175 (1980).

\bibitem{Franklin} J.~Franklin, Phys. Rev. \textbf{D66}, 033010 (2002).
 
\bibitem{pdg} W.~M.~Yao et~al. (Particle Data Group), J. Phys.
  \textbf{G33}, 1 (2006).

\bibitem{Altarelli} G.~Altarelli, G.~G.~Ross, Phys. Lett. \textbf{B212},
  391 (1988).
\bibitem{Carlitz} R.~D.~Carlitz, J.~D.~Collins, and A.~H.~Mueller,
  Phys. Lett. \textbf{B214}, 229 (1988).
\bibitem{Efremov} A.~V.~Efremov, O.~V.~Teryaev, Dubna Report No.
  JIN-E2-88-287, 1998.

\bibitem{Grasser} J.~Grasser, H.~Leutwyler, and M.~E.~Saino, Phys. Lett.
  \textbf{B253}, 252 (1991); 
  A. O. Bazarko \emph{et al.} (CCFR), Z. Phys. \textbf{C65}, 189 (1995). 

\bibitem{Theberge} S.~Theberge and A.~W.~Thomas, Phys. Rev. \textbf{D25}, 284
  (1982).
\bibitem{Cohen} J.~Cohen and H.~J. Weber, Phys. Lett. \textbf{B165}, 229 (1985).

\bibitem{NQM} G.~Karl, Phys. Rev. \textbf{D45}, 247 (1992).

\end{thebibliography}
\end{document}